\documentclass[twocolumn,aps,pra,10pt]{revtex4-1}
\usepackage{bm}
\usepackage{amsfonts}
\usepackage{amssymb}
\usepackage{amsmath}
\usepackage{graphicx}
\newcommand{\sint}{\;\int\mspace{-26mu}\sum}
\begin{document}

\title{Heisenberg uncertainty relations for photons}
\author{Iwo Bialynicki-Birula}\email{birula@cft.edu.pl}
\affiliation{Center for Theoretical Physics, Polish Academy of Sciences\\
Aleja Lotnik\'ow 32/46, 02-668 Warsaw, Poland}
\author{Zofia Bialynicka-Birula}\affiliation{Institute of Physics, Polish Academy of Sciences\\
Aleja Lotnik\'ow 32/46, 02-668 Warsaw, Poland}

\begin{abstract}
The idea to base the uncertainty relation for photons on the electromagnetic energy distribution in space enabled us to derive a sharp inequality that expresses the uncertainty relation [Phys. Rev. Lett. {\bf 108}, 140401 (2012)]. An alternative version of the uncertainty relation derived in this paper is closer in spirit to the original Heisenberg relation because it employs the analog of the position operator for the photon---the center of the energy operator. The noncommutativity of the components of the center of the energy operator results in the increase of the bound $3\hbar/2$ in the standard Heisenberg uncertainty relation in three dimensions. This difference diminishes with the increase of the photon energy. In the infinite-momentum frame, the lower bound in the Heisenberg uncertainty relations for photons is the same as in nonrelativistic quantum mechanics. A similar uncertainty relation is also derived for coherent photon beams. This relation has direct experimental consequences since it gives a precise relationship between the spectral composition of the laser beam and the minimal focal volume.
\end{abstract}

\maketitle

\section{Introduction}

The nonexistence of the strictly localized photon states \cite{nw,*aw,*tj} and the associated lack of the photon position operator makes it impossible to formulate the uncertainty relation for photons in the standard Heisenberg form. On the other hand, it is obvious that also for photons the spread of momentum and the extension in space are subjected to some restrictions that embody the famous Heisenberg phrase \cite{heis} ``Je genauer der Ort bestimmt ist, desto ungenauer ist der Impuls bekannt und umgekehrt.''

Our approach to the photon uncertainty relations is based on two precisely defined concepts: the photon wave function in momentum space and the energy density of the quantized electromagnetic field. In our previous publication \cite{prl} we used the {\em second moment} of the energy distribution to measure the spread of the photon states in coordinate space. This led us to the uncertainty relation for photons in the form
\begin{align}\label{one}
\Delta r\Delta p\ge 4\hbar.
\end{align}

In the present work, we define the uncertainty of the position for photons that would be analogous to the standard definition. For that we need some replacement for the (nonexistent) photon position operator. This role is played by ${\hat{\bm R}}$, which is the center of energy (or center of mass). The center of energy operator ${\hat{\bm R}}$ is directly related to the {\em first moment} of the energy distribution. This approach will allow us to obtain the uncertainty relation in a form even closer to the original Heisenberg relation,
\begin{align}\label{two}
\sqrt{\Delta{{\bm R}^2}}\sqrt{\Delta{{\bm P}^2}}>\frac{d}{2}\hbar,
\end{align}
where $d$ is the number of dimensions. A characteristic feature of the uncertainty relation for photons is that the left-hand side in this inequality in two and in three dimensions is never equal to $d\hbar/2$, but it tends to this limit with the increase of the average photon momentum. Only in the infinite-momentum frame is the uncertainty relation for photons the same as for nonrelativistic massive particles. However, in one dimension, the inequality (\ref{two}) is saturated so that in this case there is no difference between photons and massive nonrelativistic particles.

We also prove the following sharp inequality:
\begin{align}\label{three}
\sqrt{\langle{\hat{\bm R}}\!\cdot\!{\hat{\bm R}}\rangle}\sqrt{\langle{\hat{\bm P}}\!\cdot\!{\hat{\bm P}}\rangle}\ge\frac{3}{2}\hbar\,\sqrt{1+\frac{4\sqrt{5}}{9}}.
\end{align}

In nonrelativistic quantum mechanics, the inequalities obeyed by the two measures of uncertainty, $\Delta{{\bm R}^2}\Delta{{\bm P}^2}$ and $\langle{\hat{\bm R}}\cdot{\hat{\bm R}}\rangle\langle{\hat{\bm P}}\cdot{\hat{\bm P}\rangle}$, are completely equivalent. They have equal lower bounds and they are both saturated by Gaussian functions. This equivalence does not hold for photons. Nevertheless, the two inequalities are intimately related. We shall first prove (\ref{three}) and then use the information about the photon states that saturate this inequality to elucidate the intricate properties of the inequality (\ref{two}).

An early attempt to base the uncertainty relation for photons on the center of energy ${\hat{\bm R}}$ was made by Schwinger \cite{js}, but he only gave a rough estimate that the lower bound of $\Delta{{\bm R}^2}\Delta{{\bm P}^2}$ is of the order of $\hbar^2$.

In addition to an uncertainty relation for single photons we derive a closely related uncertainty relation for photon beams. Using coherent states of the electromagnetic field to describe such beams in the limit of a large number of photons we prove the following sharp inequality:
\begin{align}\label{four}
\sqrt{\Delta{{\bm R}^2}}\sqrt{\Delta{{\bm P}^2}}\ge\frac{3}{2}\hbar\,\sqrt{1+\frac{4\sqrt{2}}{9}},
\end{align}
and we find the mode functions of the coherent states that saturate this inequality.

\section{The center of energy}

The nonexistence of the local photon density in configuration space is due to the fact that in quantum electrodynamics the operator of the total number of photons ${\hat N}$ involves not a single but a double integral \cite{yz}:
\begin{align}\label{n1}
{\hat N}&=\frac{1}{4\pi^2\hbar c}\int\!d^3r\int\!d^3r'\nonumber\\
&\times :\!\left[\frac{{\hat{\bm D}}({\bm r},t)\!\cdot\!{\hat{\bm D}}({\bm r}',t)}{\varepsilon|{\bm r}-{\bm r}'|^2}
+\frac{{\hat{\bm B}}({\bm r},t)\!\cdot\!{\hat{\bm B}}({\bm r}',t)}{\mu|{\bm r}-{\bm r}'|^2}\right]\!:\nonumber\\
&=\frac{1}{2\pi^2\hbar c}\int\!d^3r\int\!d^3r':\!\left[\frac{{\hat{\bm F}^\dagger}({\bm r},t)\!\cdot\!{\hat{\bm F}}({\bm r}',t)}{|{\bm r}-{\bm r}'|^2}\right]\!:.
\end{align}
We use systematically the Riemann-Silberstein vector (the RS vector) \cite{pwf}
\begin{align}\label{F}
{\hat{\bm F}}(\bm r,t)=\frac{{\hat{\bm D}}(\bm r,t)}{\sqrt{2\epsilon}}+i\frac{{\hat{\bm B}}(\bm r,t)}{\sqrt{2\mu}},
\end{align}
which will allow us to write many formulas in a compact form. The normal ordering removes the (infinite) contribution from the vacuum state. In contrast to the total-number operator, the total-energy operator of the electromagnetic field ${\hat H}$ (the Hamiltonian) is an integral of a {\em local} density,
\begin{align}\label{ham}
{\hat H}=\int\!d^3r\,{\hat{\mathcal E}}({\bm r},t),
\end{align}
where
\begin{align}\label{en}
{\hat{\mathcal E}}({\bm r},t)=:\!{\hat{\bm F}^\dagger}({\bm r},t)\!\cdot\!{\hat{\bm F}}({\bm r},t)\!:.
\end{align}

The center of the energy operator can be introduced in {\em any relativistic theory}. All we need for this construction is the set of generators of the Poincar\'e group. Following Born and Infeld  \cite{bi,*pryce}, we define the operator ${\hat{\bm R}}$ as follows:
\begin{align}\label{cm}
{\hat{\bm R}}=\frac{1}{2\hat{H}}{\hat{\bm N}}+{\hat{\bm N}}\frac{1}{2\hat{H}}=\frac{1}{\sqrt{\hat{H}}}{\hat{\bm N}}\frac{1}{\sqrt{\hat{H}}},
\end{align}
where ${\hat{\bm N}}$ is the first moment of the energy distribution,
\begin{align}\label{boost}
{\hat{\bm N}}=\int\!d^3r\,{\bm r}\,{\hat{\mathcal E}}({\bm r},t).
\end{align}
The symmetrization in (\ref{cm}) is necessary to obtain a Hermitian operator. The inverse of the Hamiltonian is well defined, provided we exclude the vacuum state. The spectrum of the Hamiltonian is nonnegative, therefore the positive square root is unique. The significance of ${\hat{\bm N}}$ is further underscored by its being the generator of Lorentz transformations. Since the operators ${\hat{H}}$ and ${\hat{\bm N}}$ do not commute (the energy changes under Lorentz transformations), the equivalence of the two forms of ${\hat{\bm R}}$ in (\ref{cm}) is not obvious and is proved in Appendix A.

It follows from the commutation relations between the generators of the Poincar\'e group \cite{bi,dirac0},
\begin{align}\label{comnp}
[{\hat N}_i,{\hat P}_j]=i\hbar\delta_{ij}{\hat H},
\end{align}
that ${\hat{\bm R}}$ and the total momentum ${\hat{\bm P}}$ obey the canonical commutation relations between the position and momentum,
\begin{align}\label{comrn}
[{\hat R}_i,{\hat{P}}_j]=i\hbar\delta_{ij}.
\end{align}
We must, however, resist the temptation to treat ${\hat{\bm R}}$ as a bona fide position operator because its components {\em do not commute},
\begin{align}\label{rcom}
[{\hat R}_i,{\hat R}_j]=-i\hbar c^2{\hat H}^{-1}{\hat S}_{ij}{\hat H}^{-1},
\end{align}
where ${\hat S}_{ij}$ is the operator of the intrinsic angular momentum: the difference between the total angular momentum and the orbital angular momentum,
\begin{align}\label{spin}
{\hat S}_{ij}={\hat M}_{ij}-\left({\hat R}_i{\hat P}_j-{\hat R}_j{\hat P}_i\right).
\end{align}
Note that the effects of the noncommutativity are present in all systems with intrinsic angular momentum and decrease with the increasing energy. We shall fully confirm this observation in Sec.~\ref{ur2}.

\section{Relativistic uncertainty relations in one, two, and three dimensions}

Despite all of the differences between the nonrelativistic and relativistic dynamics we may derive a sharp Heisenberg uncertainty relation {\em along one direction}, say $x$, for any relativistic system. This one-dimensional uncertainty relation is based solely on the commutation relations between ${\hat X}={\hat{R}}_x$ and ${\hat{P}}={\hat{P}}_x$ and has the standard form
\begin{align}\label{hur}
\sqrt{\Delta X^2}\sqrt{\Delta{P}^2}\ge\textstyle\frac{1}{2}\hbar,
\end{align}
where
\begin{subequations}
\begin{align}\label{var1}
\Delta{X}^2&=\langle(\Delta\hat{P})^2\rangle,\quad
{\Delta\hat X}={\hat X}-\langle{\hat X}\rangle,\\
\Delta{P}^2&=\langle(\Delta\hat{P})^2\rangle,\quad
{\Delta\hat P}={\hat P}-\langle{\hat P}\rangle.
\end{align}
\end{subequations}
The one-dimensional uncertainty relation holds for {\em any relativistic quantum system}. A simple proof of (\ref{hur}) uses the commutation relations (\ref{comrn}) and the non-negative expectation value of the operator:
\begin{align}\label{schwartz1}
\left\langle\left(\Delta\hat{X}-i\lambda\Delta\hat{P}\right)
\left(\Delta\hat{X}+i\lambda\Delta\hat{P}\right)\right\rangle\ge 0,
\end{align}
where $\lambda$ is an arbitrary real number. The condition that this expression treated as a function of $\lambda$ can have at most one real root gives (\ref{hur}). This inequality is saturated by the quantum state whose state vector satisfies the condition
\begin{align}\label{sat1}
\left(\Delta\hat{X}+i\lambda\Delta\hat{P}\right)|\Psi\rangle=0.
\end{align}
The specific form of $|\Psi\rangle$ depends, of course, on the system under study. Note that we may remove the average values $\langle{\hat X}\rangle$ and $\langle{\hat P}\rangle$ from (\ref{sat1}) by choosing $|\Psi\rangle$ in the form
\begin{align}\label{sat2}
|\Psi\rangle=\exp\left(i\langle{\hat P}\rangle\hat{X}/\hbar-i\langle{\hat X}\rangle\hat{P}/\hbar\right)|\Psi'\rangle.
\end{align}
Since the inequality must hold for {\em all} vectors, replacing $|\Psi\rangle$ by $|\Psi'\rangle$ makes no difference and the two forms of the uncertainty relation in one dimension, namely,
\begin{align}\label{equiv}
\sqrt{\Delta{X}^2}\sqrt{\Delta{P}^2}\ge\textstyle\frac{1}{2}\hbar\quad{\text{and}} \quad\sqrt{\langle{\hat{X}^2\rangle}}\sqrt{\langle{\hat{P}^2}}\rangle
\ge\textstyle\frac{1}{2}\hbar,
\end{align}
are completely equivalent. In nonrelativistic quantum mechanics the equivalence holds  in any number of dimensions. A spherically symmetric Gaussian function shifted in the coordinate space by $\langle{\bm r}\rangle$ and in the momentum space by $\langle{\bm p}\rangle$ by the unitary transformation of the form (\ref{sat2}) will automatically saturate the inequality (\ref{two}). This equivalence, however, is no longer valid for relativistic systems in three dimensions.

To extend our analysis to two and three dimensions, we introduce the dispersion in position that involves two or three components of the center-of-energy vector ${\hat{\bm R}}$,
\begin{align}\label{r2}
\Delta{\bm R}^2=\langle{\Delta\hat{\bm R}}\!\cdot\!{\Delta\hat{\bm R}}\rangle,
\end{align}
where $\Delta\hat{\bm R}={\hat{\bm R}}-\langle{\hat{\bm R}}\rangle$ and the dispersion in momentum,
\begin{align}\label{p2}
\Delta{\bm P}^2=\langle{\Delta\hat{\bm P}}\!\cdot\!{\Delta\hat{\bm P}}\rangle,
\end{align}
where $\Delta\hat{\bm P}={\hat{\bm P}}-\langle{\hat{\bm P}}\rangle$.
Following the same procedure as the one used in deriving (\ref{hur}), we obtain (\ref{two}). The proof is based this time on the expectation value of the following positive operator:
\begin{align}\label{schwartz}
\left\langle\left(\Delta\hat{\bm R}-i\lambda\Delta\hat{\bm P}\right)\!\cdot\!\left(\Delta\hat{\bm R}+i\lambda\Delta\hat{\bm P}\right)\right\rangle > 0.
\end{align}

In contrast to the one-dimensional case, the inequalities (\ref{two}) and (\ref{schwartz}) are not sharp because there is no state vector that is annihilated by {\em all three components} of the vector operator ${\hat{\bm A}}=\Delta\hat{\bm R}+i\lambda\Delta\hat{\bm P}$ and even by two components. This is due to the fact that the commutators (\ref{rcom}) of the components of ${\hat{\bm R}}$ do not vanish. Should there exist a state vector annihilated by ${\hat A}$, then this vector would also be annihilated by the commutators of the components of ${\hat A}$. These commutators are proportional to the components of spin. Therefore, for any relativistic quantum system endowed with spin the inequality (\ref{two}) cannot be saturated.

In the next section, we introduce a convenient formalism to describe photon states that will be later applied to derive the inequalities (\ref{three}) and (\ref{four}) and also to elucidate the meaning of the inequality (\ref{two}).

\section{Quantum mechanics of photons}\label{qm}

In what follows, we shall consider one-photon states of the electromagnetic field. These states are generated from the vacuum state by the action of the photon creation operators,
\begin{align}\label{1phot}
|f\rangle=\int\!\frac{d^3k}{k}\left[f_+(\bm k)a_+^\dagger(\bm k)+f_-(\bm k)a_-^\dagger(\bm k)\right]|0\rangle,
\end{align}
where $a_\pm^\dagger(\bm k)$ create photons with momentum $\hbar{\bm k}$ and positive or negative helicity $\lambda$ (left-handed or right-handed circular polarization). We assume the normalization of these operators such that the commutation relations have the form
\begin{align}\label{cr}
\left[a_\lambda(\bm k),a_{\lambda'}^\dagger(\bm k')\right]=\delta_{\lambda\lambda'}k\,\delta^{(3)}(\bm k-\bm k').
\end{align}
This leads to the relativistic form (the volume element on the light cone $d^3k/k$ is invariant under Lorentz transformations) of the scalar product,
\begin{align}\label{sp}
\langle f^{(1)}|f^{(2)}\rangle=\sint\!\frac{1}{k}\,f_\lambda^{(1)*}(\bm k)f^{(2)}_\lambda(\bm k),
\end{align}
and the associated norm of one-photon state vectors,
\begin{align}\label{norm}
\langle f|f\rangle=||f||^2=\sint\!\frac{1}{k}\,|f_\lambda(\bm k)|^2.
\end{align}
The symbol $\displaystyle{\sint}$ stands for the summation over $\lambda$ and the integration over $\bm k$,
\begin{align}\label{sint}
\sint=\sum_\lambda\int\!d^3k.
\end{align}
The functions $f_+(\bm k)$ and $f_-(\bm k)$ are the photon wave functions in momentum space. Their moduli squared are the probability densities to find the left- or right-handed photons with momentum $\hbar{\bm k}$.

The creation and annihilation operators are connected with the field operators through the expansion of the RS operator into plane waves \cite{bb,*bb1,*bb2},
\begin{align}\label{rep}
&{\hat{\bm F}}(\bm r,t)=\sqrt{\hbar c}\int\!\frac{d^3k}{(2\pi)^{3/2}}\nonumber\\
&\times{\bm e}(\bm k)\left[a_+(\bm k)e^{i\bm k\cdot\bm r-i\omega t}+a_-^\dagger(\bm k)e^{-i\bm k\cdot\bm r+i\omega t}\right].
\end{align}
The normalized polarization vector ${\bm e}(\bm k)$ is:
\begin{align}\label{polarc}
{\bm e}({\bm k})=\frac{{\bm k}\times({\bm n}\times{\bm k})-ik({\bm n}\times{\bm k})}{\sqrt{2}\,k|{\bm n}\times{\bm k}|},
\end{align}
where ${\bm n}$ is an arbitrary unit vector.

In order to find the action of all relevant operators on one-photon states, we first express these operators in terms of creation and annihilation operators. This task is simplified by using the RS vector in the form (\ref{rep}) and we obtain \cite{bb,bb1,bb2}
\begin{subequations}\label{gens}
\begin{align}
{\hat H}&=\sint\!\frac{1}{k}\,\hbar\omega\,a_\lambda^\dagger(\bm k)a_\lambda(\bm k),\\
{\hat{\bm P}}&=\sint\!\frac{1}{k}\,\hbar{\bm k}\,a_\lambda^\dagger(\bm k)a_\lambda(\bm k),\\
{\hat{\bm M}}&=\sint\!\frac{1}{k}\,\hbar\,a_\lambda^\dagger(\bm k)\left({\bm k}\times\frac{1}{i}\bm D_\lambda+\lambda\frac{\bm k}{k}\right)a_\lambda(\bm k),\\
{\hat{\bm N}}&=\sint\!\frac{1}{k}\,\hbar\omega\,a_\lambda^\dagger(\bm k) i\bm D_\lambda a_\lambda(\bm k),
\end{align}
\end{subequations}
where $\bm D_\lambda$ is the covariant derivative in momentum space on the light cone,
\begin{align}
{\bm D}_\lambda&={\bm\nabla}-i\lambda{\bm\alpha}({\bm k}),\label{cder}\\
{\bm\alpha}({\bm k})&=i{\bm e}^*(\bm k)\!\cdot\!{\bm\nabla}{\bm e}(\bm k)=\frac{({\bm n}\!\cdot\!{\bm k})({\bm n}\times{\bm k})}{k\,|{\bm n}\times{\bm k}|^2},\label{alpha}
\end{align}
the dot denotes the scalar product of polarization vectors, and $\bm\nabla$ denotes the derivatives with respect to $\bm k$.

In relativistic quantum mechanics of photons, the generators of the Poincar\'e group  (\ref{gens}) act on the photon wave functions as follows:
\begin{subequations}\label{gensp}
\begin{align}
{\hat H}f_{\lambda}(\bm k)&=\hbar\omega\,f_{\lambda}(\bm k),\\
{\hat{\bm P}}f_{\lambda}(\bm k)&=\hbar{\bm k}\,f_{\lambda}(\bm k),\\
{\hat{\bm M}}f_{\lambda}(\bm k)&=\hbar\left({\bm k}\times\frac{1}{i}\bm D_\lambda+\lambda\frac{\bm k}{k}\right)f_{\lambda}(\bm k),\\
{\hat{\bm N}}f_{\lambda}(\bm k)&=\hbar\omega\,i\bm D_\lambda\,f_\lambda(\bm k),\label{n}
\end{align}
\end{subequations}
where we stretched our notation keeping the same symbols to denote the operators acting on the states of the field and the operators acting on the photon wave functions. Since all of these operators are Hermitian with respect to the scalar product (\ref{sp}), they generate two unitary representations $f_+(\bm k)$ and $f_-(\bm k)$ of the Poincar\'e group. These representations are concrete realizations of the general scheme described in \cite{wig,*bw}.

The center-of-energy operator ${\hat{\bm R}}$ given by the second expression in (\ref{cm}) has the following representation in quantum mechanics of photons:
\begin{align}\label{rqm}
{\hat{\bm R}}f_{\lambda}(\bm k)=i\sqrt{k}{\bm D}_\lambda\frac{1}{\sqrt{k}} f_{\lambda}(\bm k).
\end{align}
It is often convenient to replace the function $f_\lambda(\bm k)$ by its rescaled counterpart $g_\lambda(\bm k)$,
\begin{align}\label{g}
g_\lambda(\bm k)=\frac{f_\lambda(\bm k)}{\sqrt{k}}.
\end{align}
The transformation properties of $g_\lambda(\bm k)$ under the Lorentz transformations are more complicated than those of $f_\lambda(\bm k)$, but this function is similar to the nonrelativistic wave function because in contrast to (\ref{norm}) its norm (and also the scalar product) has a familiar nonrelativistic form
\begin{align}
||g||^2=\sint\!\,g_\lambda^*g_\lambda.
\end{align}
The center-of-energy operator acting on $g_\lambda(\bm k)$ is
\begin{align}\label{rqm1}
{\hat{\bm R}}g_{\lambda}(\bm k)=i{\bm D}_\lambda g_{\lambda}(\bm k).
\end{align}
As a simple application of this formula, we find now the function that saturates the general one-dimensional uncertainty relation (\ref{hur}) in the case of photons. Choosing the direction in this relation along the ${\bm n}$ vector, we find that the covariant derivative (\ref{cder}) becomes an ordinary derivative along this direction because the component of ${\bm\alpha}({\bm k})$ along ${\bm n}$ vanishes. Therefore, the function $g_\lambda(\bm k)$ which saturates the inequality is a Gaussian in the direction ${\bm n}$. This result has been obtained before by Holevo \cite{h} in the framework of estimation theory.

The extension of the Heisenberg uncertainty relation for photons from one to three dimensions is nontrivial. In the next section, we use the representation (\ref{rqm1}) of the operator ${\hat{\bm R}}$ to fulfill this aim.

\section{Uncertainty relation for the product of $\langle{\hat{\bm R}}\cdot{\hat{\bm R}}\rangle$ and $\langle{\hat{\bm P}}\cdot{\hat{\bm P}}\rangle$}\label{ur1}

The formulation of the uncertainty relation for the photon will be carried out with the use of the operators ${\hat{\bm R}}$ and ${\hat{\bm P}}$ acting on the photon wave functions $f_\lambda({\bm k})$ in momentum space. In this section we shall consider the product of the quantities $\langle{\hat{\bm R}}\cdot{\hat{\bm R}}\rangle$ and $\langle{\hat{\bm P}}\cdot{\hat{\bm P}}\rangle$, instead of their variances. The variances $\Delta{\bm R}^2$ and  $\Delta{\bm P}^2$ reduce to $\langle{\hat{\bm R}}\cdot{\hat{\bm R}}\rangle$ and $\langle{\hat{\bm P}}\cdot{\hat{\bm P}}\rangle$ only when both $\langle{\hat{\bm R}}\rangle$ and $\langle{\hat{\bm P}}\rangle$ vanish.

The quantities $\langle{\hat{\bm R}}\cdot{\hat{\bm R}}\rangle$ and $\langle{\hat{\bm P}}\cdot{\hat{\bm P}}\rangle$ expressed in terms of the rescaled wave function $g_\lambda$ are:

\begin{align}\label{rr}
\langle{\hat{\bm R}}\!\cdot\!{\hat{\bm R}}\rangle &=\frac{1}{||g||^2}\sint\!\left({\bm D}_\lambda g_\lambda\right)^*\!\!\cdot\!{\bm D}_\lambda g_\lambda\nonumber\\
&=\frac{1}{||g||^2}\sint\!\Big[{\bm \nabla}g_\lambda^*\!\cdot\!{\bm\nabla}g_\lambda+\lambda^2{\bm\alpha}^2(\bm k)g_\lambda^*g_\lambda\nonumber\\
&+i\lambda{\bm\alpha}(\bm k)\!\cdot\!
\left(g_\lambda^*{\bm\nabla}g_\lambda-g_\lambda{\bm\nabla}g_\lambda^*\right)
\Big],
\end{align}
\begin{align}
\langle{\hat{\bm P}}\!\cdot\!{\hat{\bm P}}\rangle =\frac{\hbar^2}{||g||^2}\sint\!\,g_\lambda^*{\bm k}^2g_\lambda.
\end{align}
There is one immediate conclusion that can be drawn by inspecting the integrand in the formula for $\langle{\hat{\bm R}}\cdot{\hat{\bm R}}\rangle$. Namely, the presence of ${\bm\alpha}(\bm k)$ rules out spherically symmetric functions. To obtain a finite value of $\langle{\hat{\bm R}}\cdot{\hat{\bm R}}\rangle$ we must eliminate the singularity at $|{\bm n}\times{\bm k}|=0$ by the appropriate angular dependence of $g_\lambda$. Our analytic solution will confirm this expectation. The breaking of the spherical symmetry is an important difference between the uncertainty relation for photons and for the nonrelativistic particles.

Further calculations are most easily done after the transformation of the integrals to spherical coordinates,
\begin{align}
\langle{\hat{\bm R}}\!\cdot\!{\hat{\bm R}}\rangle &=\frac{1}{||g||^2}\sum_\lambda\!\!\int_0^\infty\!\!\!\!\!dk\, k^2\!\int_0^\pi\!\!\!\!
d\theta\sin\theta\!\! \int_0^{2\pi}\!\!\!\!\!d\varphi\nonumber\\ \times&\bigg[|\partial_kg_\lambda|^2+\frac{|\partial_\theta g_\lambda|^2}{k^2}+\frac{|\partial_\varphi g_\lambda|^2}{k^2\sin^2\theta}+\frac{\lambda^2\cos^2\!\theta|g_\lambda|^2}{k^2\sin^2\!\theta}\nonumber\\
+&\frac{i\lambda\cos\theta\left(g^*_\lambda\partial_\varphi g_\lambda-g_\lambda\partial_\varphi g^*_\lambda\right)}{k^2\sin^2\!\theta}\bigg],
\end{align}
\begin{align}
\langle{\hat{\bm P}}\!\cdot\!{\hat{\bm P}}\rangle &=\frac{\hbar^2}{||g||^2}\sum_\lambda\!\!\int_0^\infty\!\!\!\!\!dk\,k^2\!\int_0^\pi\!\!\!\!
d\theta\sin\theta\!\!\int_0^{2\pi}\!\!\!\!\!d\varphi\,k^2|g_\lambda|^2,
\end{align}
\begin{align}
||g||^2 &=\sum_\lambda\!\!\int_0^\infty\!\!\!\!\!dk\,k^2\!\int_0^\pi\!\!\!\!
d\theta\sin\theta\!\! \int_0^{2\pi}\!\!\!\!\!d\varphi|g_\lambda|^2.
\end{align}
The left-hand side of the uncertainty relation for ${\hat{\bm R}}\cdot{\hat{\bm R}}$ and ${\hat{\bm P}}\cdot{\hat{\bm P}}$ divided by $\hbar^2$ is a dimensionless quantity which will be denoted \cite{gamma} by $\gamma^2$,
\begin{align}\label{gamma}
\gamma^2=\frac{\langle{\hat{\bm R}}\cdot{\hat{\bm R}}\rangle\langle{\hat{\bm P}}\cdot{\hat{\bm P}}\rangle}{\hbar^2}.
\end{align}

We shall determine the minimal value of $\gamma$ applying a variational procedure, as we have done in \cite{prl}. The variation of $\gamma^2$ with respect to $g_\lambda^*({\bm k})$ leads to the following equation for $g_\lambda({\bm k})$:
\begin{widetext}
\begin{align}\label{vareq}
\left[-\frac{1}{\kappa^2}\partial_\kappa \kappa^2\partial_\kappa-\frac{1}{\kappa^2\sin^2\theta}
\left(\partial_\theta\sin\theta\,\partial_\theta
+\partial_\varphi^2-\lambda^2-2i\lambda\cos\theta\,\partial_\varphi
\right)-\frac{\lambda^2}{\kappa^2}+\gamma\kappa^2-2\gamma\right]g_\lambda(\kappa,\theta,\varphi)=0,
\end{align}
\end{widetext}
where we replaced $k$ by the dimensionless variable $\kappa$,
\begin{align}\label{kappa}
\kappa=k\left(\hbar^2\frac{\langle{\hat{\bm R}}\cdot{\hat{\bm R}}\rangle}
{\langle{\hat{\bm P}}\cdot{\hat{\bm P}}\rangle}\right)^{1/4}.
\end{align}
After performing the variation, we put $||g||^2=1$. The variational equations for two values of $\lambda$ decouple, so that we may take one value of $\lambda$ at a time. Since the change of the sign of $\lambda$ is compensated by complex conjugation, we will consider only positive helicity $\lambda=1$.

Equation (\ref{vareq}) allows for the separation of variables,
\begin{align}\label{sep}
g_\lambda(\kappa,\theta,\varphi)={\mathcal K}(\kappa)\Theta(\theta)e^{im\varphi},
\end{align}
and we obtain the following equations for the radial and the angular parts:
\begin{align}\label{ekappa}
\left[-\frac{1}{\kappa^2}\partial_\kappa \kappa^2\partial_\kappa+\frac{j(j+1)-\lambda^2}{\kappa^2}+\gamma\kappa^2\right]{\mathcal K}(\kappa)=2\gamma{\mathcal K}(\kappa),
\end{align}
\begin{align}\label{etheta}
\left[-\frac{1}{\sin\theta}\,\partial_\theta\sin\theta\,\partial_\theta
+\frac{m^2+\lambda^2-2\lambda m\cos\theta}{\sin^2\theta}\right]
\Theta(\theta)\nonumber\\=j(j+1)\Theta(\theta).
\end{align}

The equation for $\Theta(\theta)$ is the same as in the theory of magnetic monopoles (cf. \cite{prl},\cite{mklg,*milton}). Its solutions are given in terms of Jacobi polynomials $P^{(m,m')}_j(x)$ (also known as ``monopole harmonics''),
\begin{align}\label{theta}
\Theta(\theta)=\sin^\lambda\!\theta\cot^m\!\theta P^{(\lambda-m,\lambda+m)}_{j-m}(\cos\theta).
\end{align}
Regular solutions are obtained when $j$ is a natural number starting from $j=1$. For $j=0$, not only are both solutions of the angular equation, namely, $1/\sin\theta$ and $\cot\theta$, singular, but also the radial equation does not have regular solutions because the centrifugal force becomes attractive. Therefore, the s states are ruled out as we already observed before.

The equation for ${\mathcal K}(\kappa)$ is the radial part of the Schr\"odinger equation for the three-dimensional harmonic oscillator with a modified centrifugal force. This equation after the substitution,
\begin{align}\label{subst}
{\mathcal K}(\kappa)=\kappa^{\nu-3/2}
\exp\left(-\textstyle\frac{1}{2}\kappa^2\right){\tilde{\mathcal K}}(\kappa),
\end{align}
reduces to the equation for the confluent hypergeometric function,
\begin{align}\label{conf}
{\tilde {\mathcal K}}(\kappa)=~_1F_1\left(\frac{\nu-\gamma}{2},\nu;\kappa^2\right),
\end{align}
where $\nu=1+\sqrt{j+j^2-3/4}$. To obtain a regular solution, $~_1F_1$ must become a polynomial and this leads to the quantization condition for the parameter $\gamma$,
\begin{align}\label{qc}
\gamma=2n+1+\sqrt{j+j^2-3/4},\quad n=0,1,2\dots.
\end{align}
The lowest value of $\gamma$ is obtained for $j=1$ and $n=0$,
\begin{align}\label{qc1}
\gamma=1+\frac{\sqrt{5}}{2}=\frac{3}{2}\sqrt{1+\frac{4\sqrt{5}}{9}}.
\end{align}
This is the right-hand side in the uncertainty relation (\ref{three}).

In what follows, we shall denote by $\gamma$ always its lowest value (\ref{qc1}). This eigenvalue is degenerate. There are three eigenfunctions that saturate the inequality (\ref{three}) corresponding to $m=0,\pm 1$:
\begin{subequations}\label{wfs}
\begin{align}
f_0(k,\theta,\phi)&=Aa\sin\theta\,(ak)^{\gamma-1}
\exp\left(-\textstyle\frac{1}{2}(ak)^2\right),\\
f_\pm(k,\theta,\phi)&=Aa\frac{(1\pm\cos\theta)}{\sqrt{2}}e^{\pm i\phi}
(ak)^{\gamma-1}\exp\left(-\textstyle\frac{1}{2}(ak)^2\right),
\end{align}
\end{subequations}
where the normalization constant is
\begin{align}\label{nu}
A=\sqrt{\frac{3}{4\pi\Gamma(\gamma)}},
\end{align}
and the parameter $a$ sets the length scale. The value of $a$ is arbitrary because there is no intrinsic length associated with the photon.

To exhibit the geometric structure of the wave functions (\ref{wfs}) we shall rewrite them as components of a Cartesian vector ${\bm f}=(f_x,f_y,f_z)$ in Cartesian coordinates,
\begin{align}\label{wfc}
{\bm f}({\bm k})&=Aa^2\,(ak)^{\gamma-1}
\exp\left(-\textstyle\frac{1}{2}(ak)^2\right)\nonumber\\
&\times\frac{{\bm k}\times({\bm n}\times{\bm k})+ik({\bm n}\times{\bm k})}{|{\bm n}\times{\bm k}|},
\end{align}
The presence of the unit vector in the direction ${\bm n}\times{\bm k}$ means that there is a vortex line in momentum space along the ${\bm n}$ direction with unit intensity. To obtain the formulas (\ref{wfs}) we must choose the direction of $\bm n$ as the $z$ axis in spherical coordinates.

The increase of the lower bound in the uncertainty relation (\ref{three}) from the value $3\hbar/2$ underscores the unique properties of photons. This increase is due to the specific angular dependence of the photon wave function in momentum space enforced by the nontrivial geometry on the light cone. As a result, all three functions (\ref{wfs}) vanish at ${\bm k}=0$, in contrast to the Gaussian functions saturating the standard Heisenberg relation. This effect was also present in our previous photon uncertainty relation \cite{prl}. In both cases, the angular dependence is the same. However, the radial dependence is different and this difference is reflected in the values of the lower bounds. The scaling of $\kappa$ is chosen so that (as in \cite{prl}) the uncertainties in position and momentum are equally distributed,
\begin{subequations}
\begin{align}\label{eq}
\langle{\hat{\bm R}}\cdot{\hat{\bm R}}\rangle&=a^2(1+\sqrt{5}/2),\\
\langle{\hat{\bm P}}\cdot{\hat{\bm P}}\rangle&=(\hbar/a)^2(1+\sqrt{5}/2).
\end{align}
\end{subequations}
Of course, their product is scale independent and gives the lower bound.

\section{Uncertainty relation for the product of $\Delta{\bm R}^2$ and $\Delta{\bm P}^2$}\label{ur2}

The information gained in the analysis of the uncertainty relation (\ref{three}) will now be used to improve the bound in (\ref{two}). The first observation is that $\langle{\hat{\bm R}}\rangle=0$ for all three functions that saturate (\ref{three}), whereas the value of $\langle{\hat{\bm P}}\rangle$ does not vanish for the states with $m=\pm1$,
\begin{align}\label{avp}
\langle{\hat{\bm P}}\rangle=\pm\frac{\Gamma(3/2+\sqrt{5}/2)}{2\Gamma(1+\sqrt{5}/2)}\frac{\hbar}{a}\,{\bm n}=\pm 0.686\frac{\hbar}{a}\,{\bm n}.
\end{align}
Thus, already in this simple case the value of $\Delta{{\bm R}^2}\Delta{{\bm P}^2}=\langle{\hat{\bm R}}\cdot{\hat{\bm R}}\rangle\langle{\hat{\bm P}}\cdot{\hat{\bm P}}\rangle-\langle{\hat{\bm R}}\cdot{\hat{\bm R}}\rangle\langle{\hat{\bm P}}\rangle^2$ is lower than the value of $\langle{\hat{\bm R}}\cdot{\hat{\bm R}}\rangle\langle{\hat{\bm P}}\cdot{\hat{\bm P}}\rangle$.

\begin{figure}\label{fig1}
\includegraphics[scale=0.95]{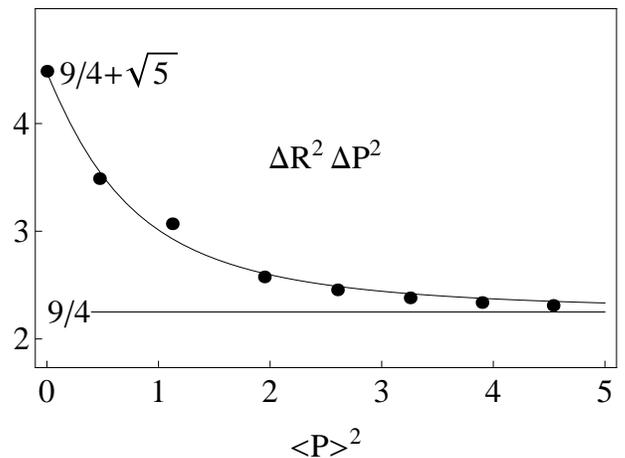}
\caption{Dependence of the product of variances (in units of $\hbar^2$) on the squared mean momentum (in units of $\hbar^2/a^2$). The leftmost dot represents the exact value of $\gamma^2$ obtained for vanishing average momentum. The remaining dots (from left to right) mark the values obtained for the trial functions (\ref{pol}) with none, one, two, and up to six terms. The solid curve represents the two-parameter fit (\ref{fit}).}
\end{figure}
In the general case, the bigger $\langle{\hat{\bm P}}\rangle$ is, the bigger will be the average photon energy. Thus, the noncommutativity of the components of ${\hat{\bm R}}$ plays a decreasing role, bringing us closer to the situation in nonrelativistic quantum mechanics. This is clearly seen in Fig.~1 where we show the exact value (\ref{three}) obtained for the vanishing mean momentum and the results of numerical calculations of $\Delta{{\bm R}^2}\Delta{{\bm P}^2}$. The points in this plot were obtained by choosing the trial functions as the product of $f_\pm(k,\theta,\phi)$ and a polynomial in $k\cos\theta$,
\begin{align}\label{pol}
1+a_1k\cos\theta+a_2(k\cos\theta)^2+a_3(k\cos\theta)^3\cdots,
\end{align}
where $a_i$ are variational parameters. These parameters are determined by requiring that they give the lowest value of $\Delta{{\bm R}^2}\Delta{{\bm P}^2}$. The points in Fig.~1 represent the values obtained with none, one, two, and up to six parameters. The solid line represents a simple two-parameter fit of the form
\begin{align}\label{fit}
9/4+\frac{\sqrt{5}}{1+1.14\,\langle{\hat{\bm P}}\rangle^{2}+0.8\,\langle{\hat{\bm P}}\rangle^{4}},
\end{align}
to all eight results. The numerical results clearly show the convergence to the value 9/4 when $\langle{\hat{\bm P}}\rangle$ tends to infinity. This result is also in agreement with the formula (\ref{rcom}) for the commutator of the center-of-energy operators since the right-hand side tends to zero with the increase of the energy, so that at infinite energy these operators behave as their nonrelativistic counterparts. We shall confirm now this result with analytic considerations. We show that in the infinite-momentum frame, we indeed obtain as a lower bound in the uncertainty relation the limiting value $3\hbar/2$. Thus, our aim is to find the minimal value of the expression
\begin{align}\label{min}
\Delta{\bm R}^2\Delta{\bm P}^2=\left\langle\left({\hat{\bm R}}-\langle{\hat{\bm R}}\rangle\right)^2\right\rangle\left\langle\left({\hat{\bm P}}-\langle{\hat{\bm P}}\rangle\right)^2\right\rangle,
\end{align}
in the limit of infinite $\langle{\hat{\bm P}}\rangle$.

In the first step, we eliminate $\langle{\hat{\bm R}}\rangle$ by applying the unitary transformation $\exp(-i\langle{\hat{\bm R}}\rangle\cdot{\hat{\bm P}}/\hbar)$ (i.e., by choosing the center of the energy as the origin of the coordinate system). The elimination of $\langle{\hat{\bm P}}\rangle$ by the unitary transformation $\exp(i\langle{\hat{\bm P}}\rangle\cdot{\hat{\bm R}}/\hbar)$ is not so painless because the components of ${\hat{\bm R}}$ do not commute and we are left with the expression
\begin{align}\label{min1}
\gamma^2=\frac{1}{\hbar^2}\left\langle e^{-i\langle{\hat{\bm P}}\rangle\cdot\hat{\bm R}/\hbar}\,{\hat{\bm R}}\!\cdot\!{\hat{\bm R}}\,e^{i\langle{\hat{\bm P}}\rangle\cdot\hat{\bm R}/\hbar}\right\rangle\left\langle{\hat{\bm P}}\!\cdot\!{\hat{\bm P}}\right\rangle,
\end{align}
which is to be minimized \cite{gamma}. In Appendix B we find by the variational procedure that the minimum of $\gamma$ is indeed equal to $3\hbar/2$.

The significant simplification of relativistic dynamics in the infinite momentum frame was noted a long time ago \cite{wein}. In particular, it has been shown \cite{suss,*ks} that in this limit, the symmetry group in the transverse plane is the Galilean group in two dimensions that governs nonrelativistic quantum mechanics. This explains why ${\bm\alpha}({\bm k})$, given in the infinite-momentum frame by (\ref{lima}), is very simple leading to the nonrelativistic lower bound in the uncertainty relation for photons (\ref{two}).

\section{Uncertainty relation for photon beams}\label{beams}

In most experiments photons appear in the form of photon beams. In this section we derive the uncertainty relation for a very common representation of such beams: the coherent state of the electromagnetic field. The exact determination of the uncertainty relation for the coherent state does not seem to be feasible but the important case---the limit when the mean photon number $\langle N\rangle$ is large---is tractable.

Coherent states $|\text{coh}\rangle$ are generated from the vacuum state by the unitary Glauber displacement operator $D$ \cite{rg1},
\begin{align}\label{coh}
&D=\exp\left(\!\sqrt{\langle N\rangle}\sint\!\frac{1}{k}\left[f_\lambda(\bm k)a_\lambda^\dagger(\bm k)-f_\lambda^*(\bm k)a_\lambda(\bm k)\right]\right),\nonumber\\
&|{\text{coh}}\rangle=D|0\rangle,
\end{align}
where the function $f_\lambda(\bm k)$ that so far represented a single-photon state now describes an arbitrary nonmonochromatic mode of electromagnetic radiation \cite{tg,*sr}. We pulled out the square root of the mean photon number ${\langle N\rangle}$ in the coherent state to have better control of the large ${\langle N\rangle}$ limit. The function $f$ will be normalized to 1 as in (\ref{norm}).

Our aim, as in Sec.~\ref{ur2}, is to minimize the left-hand side of the uncertainty relation (\ref{min}). This time \cite{gamma}, all expectation values are to be evaluated in the coherent state (\ref{coh}),
\begin{align}\label{exp1}
\gamma^2=\frac{\left(\langle{\hat{\bm R}}\!\cdot\!{\hat{\bm R}}\rangle-\langle{\hat{\bm R}}\rangle\!\cdot\!\langle{\hat{\bm R}}\rangle\right)\left(\langle{\hat{\bm P}}\!\cdot\!{\hat{\bm P}}\rangle-\langle{\hat{\bm P}}\rangle\!\cdot\!\langle{\hat{\bm P}}\rangle\right)}{\hbar^2}.
\end{align}
A fairly complicated evaluation of the two factors appearing in this formula is relegated to Appendix C. Using the formulas (\ref{fin3}) and (\ref{fin2}) we obtain the following expression for $\gamma^2$ valid for large values of ${\langle N\rangle}$:
\begin{equation}\label{gam3}
\gamma^2=\left.\begin{aligned}\underline{\sint\!k|{\bm D}_\lambda}&
\underline{f_\lambda(\bm k)|^2\sint\!k|f_\lambda(\bm k)|^2}\\
&\Big[\sint\!|f_\lambda(\bm k)|^2\Big]^{2}\end{aligned}\right.+{\mathcal O}\left(\frac{1}{\langle N\rangle}\right).
\end{equation}
In what follows we will tacitly assume that all results are valid only in the limit when ${\langle N\rangle}\to\infty$, and we will omit the symbol ${\mathcal O}(1/{\langle N\rangle})$.

Before subjecting this expression to the variational procedure, let us note that it does not depend on the normalization of $f_\lambda(\bm k)$. Therefore, we may vary the function $f_\lambda(\bm k)$ freely, as we did in all previous cases. The variation with respect to $f_\lambda^*(\bm k)$ leads to the following equation for $f_\lambda$ in the spherical coordinate system:
\begin{widetext}
\begin{align}\label{vareq4}
\left[-\frac{1}{\kappa^3}\partial_\kappa \kappa^3\partial_\kappa-\frac{1}{\kappa^2\sin^2\theta}
\left(\partial_\theta\sin\theta\,\partial_\theta
+\partial_\varphi^2-\lambda^2-2i\lambda\cos\theta\,\partial_\varphi
\right)-\frac{\lambda^2}{\kappa^2}-\frac{2\gamma^2}{\kappa}+\gamma^2\right]f_\lambda(\kappa,\theta,\varphi)=0.
\end{align}
\end{widetext}
We omitted here all intermediate steps because they are analogous to those followed in Sec.~\ref{ur1}. The dimensionless parameter $\kappa$ is defined now as
\begin{equation}\label{kap3}
\kappa=\frac{\hbar k}{c}\frac{\langle{\hat H}\rangle}{\Delta{{\bm P}^2}}=k\left.\begin{aligned}&\underline{\;\;\sint\!|f_\lambda(\bm k)|^2\;\;}\\
&\sint\!k|f_\lambda(\bm k)|^2\end{aligned}\right..
\end{equation}
After the separation of variable we obtain the following equation for the radial part:
\begin{align}\label{ekappa2}
\left[-\frac{1}{\kappa^3}\partial_\kappa \kappa^3\partial_\kappa+\frac{j(j+1)-\lambda^2}{\kappa^2}-\frac{2\gamma^2}{\kappa}\right]{\mathcal K}(\kappa)=-\gamma^2{\mathcal K}(\kappa),
\end{align}
while the angular part is the same as in (\ref{etheta}), so that the lowest allowed value of $j$ is 1. The equation for the radial part after the substitution,
\begin{align}\label{subst1}
{\mathcal K}(\kappa)=\kappa^{\sqrt{j(j+1)}-1}
\exp\left(-\gamma\kappa\right){\tilde{\mathcal K}}(\kappa),
\end{align}
reduces to the equation for the confluent hypergeometric function,
\begin{align}\label{conf1}
{\tilde {\mathcal K}}(\kappa)=~_1F_1\left(\frac{\mu}{2}-\gamma,\mu;2\gamma\kappa\right),
\end{align}
where $\mu=1+2\sqrt{j(j+1)}$. To obtain a regular solution, $~_1F_1$ must become a polynomial and this leads to the quantization condition for the parameter $\gamma$,
\begin{align}\label{qc2}
\gamma=\frac{2n+1+2\sqrt{j(j+1)}}{2},\quad n=0,1,2\dots.
\end{align}
The lowest value of $\gamma$ is obtained for $j=1$ and $n=0$,
\begin{align}\label{qc3}
\gamma=\frac{1}{2}+\sqrt{2}=\frac{3}{2}\sqrt{1+\frac{4\sqrt{2}}{9}}.
\end{align}
Again, as in Sec.~\ref{ur2}, the solution corresponding to the lowest value of $\gamma$ has a threefold degeneracy. The three normalized solutions, which are the counterparts of (\ref{wfs}), are
\begin{subequations}\label{wfs1}
\begin{align}
f_0(k,\theta,\phi)&=Aa\sin\theta\,(ak)^{\sqrt{2}-1}e^{-\gamma ak},\\
f_\pm(k,\theta,\phi)&=Aa\frac{(1\pm\cos\theta)}{\sqrt{2}}e^{\pm i\phi}
(ak)^{\sqrt{2}-1}e^{-\gamma ak},
\end{align}
\end{subequations}
where
\begin{align}\label{nc1}
A=(2\gamma)^{\sqrt{2}}\sqrt{\frac{3}{8\pi\Gamma(2\sqrt{2})}},
\end{align}
and the parameter $a$ sets the scale as in the case of a single photon.

\section{Observable consequences of uncertainty relations}

Our uncertainty relations for individual photons can be connected with observations through the Glauber theory of photodetection \cite{rg,*mw}, as we have indicated in \cite{prl}. The interpretation of the uncertainty relation for photons is basically the same as in the case of the standard Heisenberg uncertainty relation. The only difference is that the photodetection relies on the energy density of photons---the photon is where its energy is localized---rather than on the probability density to find the particle (its charge or mass) at a given location. To test our uncertainty relation, one would have to make repeated measurements on photons produced by the same source.

The uncertainty relation plays a different role in the case of photon beams. In this case, the limitation on the dispersion $\Delta{\bm R}^2$ imposed by the uncertainty relation finds its physical interpretation in terms of the directly observable quantity: the focal volume. Of course, the focal volume does not have sharp boundaries. However, the moments of the energy distribution give reasonable measures of its size. Thus, a sensible measure of the size of the focal volume $V_f$ is:
\begin{align}\label{fv}
V_f=\left(\Delta{{\bm R}^2}\right)^{3/2}.
\end{align}
The uncertainty relation in three dimensions gives precise bounds on the size of the focal volume for a given spectral composition of the beam. According to this relation, the decrease of $V_f$ is limited by the dispersion of momentum:
\begin{align}\label{fv1}
V_f\ge\frac{\hbar^3\gamma^3}{\left(\Delta{{\bm P}^2}\right)^{3/2}}.
\end{align}
It is worth mentioning here that the one-dimensional uncertainty relation (\ref{hur}) can give only a rough estimate of the focal volume due to the strong correlations imposed by the noncommutativity of the components of ${\hat{\bm R}}$.

The reduction of the size of the focal volume is important in many practical applications of laser beams, such as fluorescence microscopy, optical tweezers, material processing and also in medicine. We are far from suggesting that our uncertainty relations will lead to an improvement in any of these techniques, but we believe that they are relevant at the fundamental level.

\section{Conclusions}

In this work, we based the uncertainty relation for photons on a measure of the spatial extension of the photon wave function, which is built around the center-of-energy vector: the first moment of the energy distribution divided by the total energy. By replacing the second moment of energy used in Ref.~\cite{prl} by the first moment of energy, we were able to bring the analysis closer to the standard quantum-mechanical treatment.

The center-of-energy vector turned out to be a very good substitute for the nonexistent photon position operator, although the noncommutativity of its components leads to significant differences compared to the nonrelativistic case. In nonrelativistic Heisenberg uncertainty relations, the lowest value of $\sqrt{\Delta{{\bm R}^2}}\sqrt{\Delta{{\bm P}^2}}$ does not depend on the average position and on the average momentum. It is not so for photons. The lowest possible value of $\sqrt{\Delta{{\bm R}^2}}\sqrt{\Delta{{\bm P}^2}}$ depends on the choice of the Lorentz frame. It varies between $3/2\,\hbar(1+4\sqrt{5}/9)$ and $3\hbar/2$, when the average momentum changes from 0 to infinity. Somewhat paradoxically, highly energetic photons obey almost the same uncertainty relations as nonrelativistic particles. This is explained by the special properties of relativistic dynamics in the infinite-momentum frame.

The uncertainty relations based on the center-of-energy operator were also derived for photon beams described by coherent states of the electromagnetic field. Analytic results were obtained in the limit of a large number of photons in the beam. These uncertainty relation give a fundamental limitation on the reduction of the beam focal volume.

\acknowledgments

We thank the anonymous referee for the insistence that we include the discussion of the experimental consequences which made this work more complete. We also thank {\L}ukasz Rudnicki for helpful comments. This research was partly supported by the grant from the Polish Ministry of Science and Higher Education for the years 2010--2012.

\appendix

\section{}

To prove the equality of the two forms of ${\hat{\bm R}}$ in (\ref{cm}) we will first prove the following lemma:
\begin{align}\label{lem}
{\text{If}}\quad[{\hat H},{\hat C}]=0\quad{\text{then}}\quad[\sqrt{\hat H},{\hat C}]=0.
\end{align}
In the proof, we use the fact that the eigenvectors of the Hamiltonian form a basis. Acting on an arbitrary state in this basis $|E\rangle$ (excluding the vacuum), we have
\begin{align}\label{lem1}
\left(\sqrt{\hat H}+\sqrt{E}\right)[\sqrt{\hat H},{\hat C}]|E\rangle=[{\hat{H}},{\hat{C}}]|E\rangle=0.
\end{align}
Since the factor $\left(\sqrt{\hat H}+\sqrt{E}\right)$ does not vanish, it can be dropped and the validity of the lemma is established.

Next, we use the commutation relations between the Hamiltonian and the generator of the Lorentz transformations
\begin{align}\label{crel}
[{\hat H},{\hat{\bm N}}]=-i\hbar{\hat{\bm P}},
\end{align}
to obtain
\begin{align}\label{diff}
&\left[{\hat{H}},\left[\frac{1}{\sqrt{\hat{H}}}{\hat{\bm N}}\frac{1}{\sqrt{\hat{H}}},\frac{1}{\sqrt{\hat{H}}}\right]\right]\nonumber\\
=&\left[\frac{1}{\sqrt{\hat{H}}}\left[{\hat{H}},{\hat{\bm N}}\right]\frac{1}{\sqrt{\hat{H}}},\frac{1}{\sqrt{\hat{H}}}\right]
=\frac{\hbar}{i}\left[\frac{{\hat{\bm P}}}{\hat{H}},\frac{1}{\sqrt{\hat{H}}}\right]=0.
\end{align}
Finally, using the lemma, we may replace ${\hat H}$ by $\sqrt{\hat H}$ in the first term and expand the resulting double commutator:
\begin{align}\label{fin}
&0=\left[{\sqrt{\hat{H}}},\left[\frac{1}{\sqrt{\hat{H}}}{\hat{\bm N}}\frac{1}{\sqrt{\hat{H}}},\frac{1}{\sqrt{\hat{H}}}\right]\right]\nonumber\\
&=\frac{1}{\hat{H}}{\hat{\bm N}}+{\hat{\bm N}}\frac{1}{\hat{H}}-2\frac{1}{\sqrt{\hat{H}}}{\hat{\bm N}}\frac{1}{\sqrt{\hat{H}}}.
\end{align}
The vanishing of the difference of two expressions for ${\hat{\bm R}}$ appearing in (\ref{cm}) means that they are equal.

\section{}

To apply the variational procedure, we rewrite the functional (\ref{min1}) in the one-photon space. To simplify the calculations, we choose ${\bm n}$ in the direction of the average momentum. With this choice, the operator $i{\bm n}\!\cdot\!{\hat{\bm R}}$ reduces to an ordinary derivative with respect to $k_z$ because the scalar product ${\bm n}\!\cdot\!{\bm\alpha}$ vanishes. Therefore, the unitary operator $e^{i\langle{\hat{\bm P}}\rangle\cdot\hat{\bm R}/\hbar}$ acting on the photon wave functions becomes just the shift operator. Therefore, in the functional (\ref{rr}) the argument $k_z$ of ${\bm\alpha}({\bm k})$ is shifted now by $\langle k_z\rangle=\langle{\hat{P}_z}/\hbar\rangle$. The solution of the differential equation obtained by varying $\gamma^2$ is a very difficult task because the variables $k$ and $\theta$ can no longer be separated.

However, in the limiting case when $\langle k_z\rangle$ tends to infinity, there is a radical simplification. In this limit, ${\bm\alpha}({\bm k}+\langle{\bm k}\rangle)$ becomes
\begin{align}\label{lima}
\lim_{\langle{\bm k}\rangle\to\infty}{\bm\alpha}({\bm k}+\langle{\bm k}\rangle)=\frac{{\bm n}\times{\bm k}}{|{\bm n}\times{\bm k}|^2},
\end{align}
and the variational equation (\ref{vareq}) is replaced now by
\begin{widetext}
\begin{align}\label{vareq1}
\left[-\frac{1}{\kappa^2}\partial_\kappa \kappa^2\partial_\kappa-\frac{1}{\kappa^2\sin^2\theta}
\left(\partial_\theta\sin\theta\,\partial_\theta
+\partial_\varphi^2-\lambda^2-2i\lambda\,\partial_\varphi
\right)+\kappa^2-2\gamma\right]g_\lambda(\kappa,\theta,\varphi)=0,
\end{align}
\end{widetext}
which again allows for the separation of variables,
\begin{align}\label{sep1}
g_\lambda(\kappa,\theta,\varphi)={\mathcal K}(\kappa)\Theta(\theta)e^{im\varphi}.
\end{align}
The radial and the angular parts satisfy the equations
\begin{align}\label{ekappa1}
\left[-\frac{1}{\kappa^2}\partial_\kappa \kappa^2\partial_\kappa+\frac{j(j+1)}{\kappa^2}+\kappa^2\right]{\mathcal K}(\kappa)=2{\gamma}{\mathcal K}(\kappa),
\end{align}
\begin{align}\label{etheta1}
\left[-\frac{1}{\sin\theta}\,\partial_\theta\sin\theta\,\partial_\theta
+\frac{(m-\lambda)^2}{\sin^2\theta}\right]
\Theta(\theta)=j(j+1)\Theta(\theta).
\end{align}
This time, the value $j=0$ is allowed provided we choose $m=\lambda$. The equation for the radial part is that of the spherically symmetric harmonic oscillator. The lowest value of $\gamma=3\hbar/2$ is obtained for the ground state. This confirms the prediction made on the basis of our numerical calculations.

\section{}

In this appendix, we evaluate the leading terms of the expansion in $1/{\langle N\rangle}$ for the dispersion in position (\ref{r2}) and momentum (\ref{p2}) evaluated in the coherent state of the electromagnetic field. In this calculation, we use the second form of the position operator (\ref{cm}). The expectation value of any combination of creation and annihilation operators in a coherent state is tantamount to the vacuum expectation value of the same combination of these operators transformed by the action of the displacement operator $D$,
\begin{subequations}
\begin{align}\label{trans}
D^\dagger a^\dagger_\lambda(\bm k)D=a^\dagger_\lambda(\bm k)+\sqrt{\langle N\rangle}f_\lambda^*(\bm k),\\
D^\dagger a_\lambda(\bm k)D=a_\lambda(\bm k)+\sqrt{\langle N\rangle}f_\lambda(\bm k).
\end{align}
\end{subequations}
We will need only the following lowest-order correction to the operators ${\hat{H}}$, ${\hat{\bm N}}$, and ${\hat{\bm P}}$:
\begin{widetext}
\begin{subequations}
\begin{align}\label{corr}
D^\dagger{\hat{H}}D&={\langle N\rangle}\hbar c\left[\sint\!f_\lambda^\dagger(\bm k) f_\lambda(\bm k)+\frac{1}{\sqrt{\langle N\rangle}}\sint\!\left(a^\dagger_\lambda(\bm k) f_\lambda(\bm k)+f^*_\lambda(\bm k) a_\lambda(\bm k)\right)+{\mathcal O}(\frac{1}{\langle N\rangle})\right],\\
D^\dagger {\hat{\bm N}}D&={\langle N\rangle}\hbar c\left[\sint\!f_\lambda^\dagger(\bm k) i{\bm D}_\lambda f_\lambda(\bm k)+\frac{1}{\sqrt{\langle N\rangle}}\sint\!\left(a^\dagger_\lambda(\bm k)i{\bm D}_\lambda f_\lambda(\bm k)+f^*_\lambda(\bm k)i{\bm D}_\lambda a_\lambda(\bm k)\right)+{\mathcal O}(\frac{1}{\langle N\rangle})\right],\\
D^\dagger {\hat{\bm P}}D&={\langle N\rangle}\hbar\left[\sint\!f_\lambda^\dagger(\bm k){\bm n} f_\lambda(\bm n)+\frac{1}{\sqrt{\langle N\rangle}}\sint\!\left(a^\dagger_\lambda(\bm k){\bm n}f_\lambda(\bm k)+f^*_\lambda(\bm k){\bm n}a_\lambda(\bm k)\right)+{\mathcal O}(\frac{1}{\langle N\rangle})\right].
\end{align}
\end{subequations}

The first two formulas lead to the following expression for ${\hat{\bm R}}$:
\begin{align}\label{corr1}
D^\dagger {\hat{\bm R}}D&=\frac{1}{\mathcal H}\bigg[{\bm{\mathcal N}}+\frac{1}{\sqrt{\langle N\rangle}}\sint\!\left(a^\dagger_\lambda(\bm k)i{\bm D}_\lambda f_\lambda(\bm k)+f^*_\lambda(\bm k)i{\bm D}_\lambda a_\lambda(\bm k)\right)\nonumber\\
&-\frac{\bm{\mathcal R}}{\sqrt{\langle N\rangle}}\sint\!\left(a^\dagger_\lambda(\bm k) f_\lambda(\bm k)+f^*_\lambda(\bm k) a_\lambda(\bm k)\right)+{\mathcal O}(\frac{1}{\langle N\rangle})\bigg],
\end{align}
\end{widetext}
where
\begin{subequations}
\begin{align}\label{h0}
{\mathcal H}&=\sint\!\,f^*_\lambda(\bm k) f_\lambda(\bm k),\\
{\bm{\mathcal N}}&=\sint\!\,f^*_\lambda(\bm k)i{\bm D}_\lambda f_\lambda(\bm k)\\
{\bm{\mathcal R}}&={\bm{\mathcal N}}/{\mathcal H}.
\end{align}
\end{subequations}
In both factors of (\ref{exp1}), the leading terms cancel because they are c numbers, so that there is no difference between the averaged square and the square of the average. We shall first calculate the next-order corrections to the difference $\langle{\hat{\bm R}}\!\cdot\!{\hat{\bm R}}\rangle-\langle{\hat{\bm R}}\rangle\!\cdot\!\langle{\hat{\bm R}}\rangle$. First, note that if the contribution comes from only one ${\hat{\bm R}}$, then it does not contribute to the difference because it cancels out between the two terms. The ${\mathcal O}(1/\langle N\rangle)$ terms are not canceled by their counterparts in $\langle{\hat{\bm R}}\rangle\!\cdot\!\langle{\hat{\bm R}}\rangle$ only when the corrections appear in both operators ${\hat{\bm R}}$ in ${\hat{\bm R}}\!\cdot\!{\hat{\bm R}}$. The same observation holds for the momentum operator. Therefore, the lowest-order corrections come only from the products of two terms linear in the creation and annihilation operators, and the final results can be written in the form
\begin{align}\label{fin1}
&\left\langle\left({\hat{\bm R}}-\langle{\hat{\bm R}}\rangle\right)^2\right\rangle=\frac{1}{{\mathcal H}^2\langle N\rangle}\nonumber\\
&\times\sint k\left[(i{\bm D}_\lambda -{\bm{\mathcal R}})f_\lambda(\bm k)\right]^*\!\cdot\!(i{\bm D}_\lambda -{\bm{\mathcal R}})f_\lambda(\bm k),\\
\label{fin2}
&\left\langle\left({\hat{\bm P}}-\langle{\hat{\bm P}}\rangle\right)^2\right\rangle
={\langle N\rangle}\hbar^2\sint\!k f_\lambda^*(\bm k)f_\lambda(\bm k).
\end{align}
Without any loss of generality [the function $f_\lambda(\bm k)$ is at this point arbitrary and it will be determined from the variational procedure later], we can make the following replacement:
\begin{align}\label{repl}
f_\lambda(\bm k) \to \exp(-i{\bm k}\!\cdot\!{\bm{\mathcal R}})f_\lambda(\bm k).
\end{align}
This change of phase makes no difference in (\ref{fin2}), but it leads to the elimination of the ${\bm{\mathcal R}}$-dependent terms in (\ref{fin1}), and we obtain
\begin{align}\label{fin3}
\left\langle\left({\hat{\bm R}}-\langle{\hat{\bm R}}\rangle\right)^2\right\rangle=\frac{1}{{\mathcal H}^2\langle N\rangle}\sint\!k|{\bm D}_\lambda f_\lambda(\bm k)|^2.
\end{align}


\begin{thebibliography}{27}
\bibitem{gamma} The same symbol $\gamma$ will be used in this paper in three different version of the uncertainty relation. We hope that this will not lead to confusion.
\end{thebibliography}
\end{document}